\documentclass[aps,prd,twocolumn,amssymb,eqsecnum,showpacs,showkeyes,secnumarabic,graphics,floatfix,nofootinbib,tightenlines,longbibliography]{revtex4-1}
\usepackage{graphicx}
\usepackage{bm}
\usepackage{cancel}
\usepackage{epsfig}
\usepackage{dcolumn}
\usepackage{amsmath}
\usepackage{hhline}
\usepackage{enumerate}
\usepackage[utf8]{inputenc}
\usepackage[dvipsnames]{xcolor}
\usepackage[breaklinks=true,colorlinks=true,
linkcolor=blue,urlcolor=Blue,citecolor=MidnightBlue,
bookmarks=true,bookmarksopenlevel=2]{hyperref}
\usepackage{bm}
\usepackage{caption}
\usepackage{amsmath}
\usepackage {amssymb}
\usepackage{subcaption}

\def \beq  {\begin{equation}}
\def \eeq  {\end{equation}}
\def \ber  {\begin{eqnarray}}
\def \eer  {\end{eqnarray}}

\def \om    {\Omega}

\def \om0m {\Omega_{0\rm m}}

\begin{document}
\newcommand{\newc}{\newcommand}

\newc{\be}{\begin{equation}}
\newc{\ee}{\end{equation}}
\newc{\ba}{\begin{eqnarray}}
\newc{\ea}{\end{eqnarray}}
\newc{\bea}{\begin{eqnarray}}
\newc{\eea}{\end{eqnarray}}
\newc{\D}{\partial}
\newc{\ie}{{\it i.e.} }
\newc{\eg}{{\it e.g.} }
\newc{\etc}{{\it etc.} }
\newc{\etal}{{\it et al.}}
\newc{\lcdm}{$\Lambda$CDM }
\newc{\lcdmnospace}{$\Lambda$CDM}
\newc{\wcdm}{$w$CDM }
\newc{\plcdm}{Planck/$\Lambda$CDM }
\newc{\plcdmnospace}{Planck/$\Lambda$CDM}
\newc{\wlcdm}{WMAP7/$\Lambda$CDM }
\newc{\wlcdmnospace}{WMAP7/$\Lambda$CDM}
\newcommand{\fs}{{\rm{\it f\sigma}}_8}

\newcommand{\nn}{\nonumber}
\newc{\ra}{\Rightarrow}
\title{$H_0$ Tension, Phantom Dark Energy and Cosmological Parameter Degeneracies}

\author{G. Alestas}\email{g.alestas@uoi.gr}
\affiliation{Department of Physics, University of Ioannina, GR-45110, Ioannina, Greece}
\author{L. Kazantzidis}\email{l.kazantzidis@uoi.gr}
\affiliation{Department of Physics, University of Ioannina, GR-45110, Ioannina, Greece}
\author{L. Perivolaropoulos}\email{leandros@uoi.gr}
\affiliation{Department of Physics, University of Ioannina, GR-45110, Ioannina, Greece}

\date{\today}
\begin{abstract}
Phantom dark energy ($w<-1$) can produce amplified cosmic acceleration at late times, thus increasing the value of $H_0$ favored by CMB data and releasing the tension with local measurements of $H_0$. We show that the best fit value of $H_0$ in the context of the CMB power spectrum is degenerate with a constant equation of state parameter $w$,  in accordance with the approximate effective linear equation $H_0 + 30.93\; w - 36.47 = 0$  ($H_0$ in $km \; sec^{-1} \; Mpc^{-1}$). This equation is derived by assuming that both $\Omega_{0 \rm m}h^2$ and $d_A=\int_0^{z_{rec}}\frac{dz}{H(z)}$ remain constant (for invariant CMB spectrum) and equal to their best fit Planck/$\Lambda$CDM values as $H_0$, $\Omega_{0 \rm m}$ and $w$ vary.  For $w=-1$, this linear degeneracy equation leads to the best fit  $H_0=67.4 \; km \; sec^{-1} \; Mpc^{-1}$ as expected. For $w=-1.22$ the corresponding predicted CMB best fit Hubble constant is  $H_0=74 \; km \; sec^{-1} \; Mpc^{-1}$  which is identical with the value obtained  by local distance ladder measurements while the best fit matter density parameter is predicted to decrease since $\Omega_{0 \rm m}h^2$ is fixed.  We verify the above $H_0-w$ degeneracy equation by fitting a $w$CDM model with fixed values of $w$ to the Planck TT spectrum showing also that the quality of fit ($\chi^2$) is similar to that of \lcdmnospace. However, when including SnIa, BAO or growth data the quality of fit becomes worse than \lcdm when $w< -1$. Finally, we generalize the $H_0-w(z)$ degeneracy equation for the parametrization $w(z)=w_0+w_1\;  z/(1+z)$ and identify analytically the full $w_0-w_1$ parameter region (straight line) that leads to a best fit  $H_0=74\; km \; sec^{-1} \;  Mpc^{-1}$ in the context of the Planck CMB spectrum. This exploitation of $H_0-w(z)$ degeneracy can lead to immediate identification of all parameter values of a given $w(z)$ parametrization that can potentially resolve the $H_0$ tension.
 
\end{abstract}

\maketitle
\section{Introduction}
The discrepancy in the value of the Hubble parameter as obtained  from Cosmic Microwave Background (CMB) and Baryon Acoustic Oscillation (BAO) data ($67.4 \pm 0.5 \; km \; sec^{-1} \; Mpc^{-1}$ ) \cite{Ade:2015xua,Aghanim:2018eyx} and local distance ladder measurements ($H_0=74.03 \pm 1.42\;km \, s^{-1}\, Mpc^{-1}$) \cite{Riess:2019cxk} has reached a level close to $6\sigma$ \cite{Riess:2020sih, Kenworthy:2019qwq, Wong:2019kwg, Verde:2019ivm, Taubenberger:2019qna} and is becoming a problem of the standard \lcdm model. A similar issue, with lower significance level, appears when measuring the growth rate of cosmological perturbations using peculiar velocities (Redshift Space Distortions) \cite{Macaulay:2013swa,Johnson:2015aaa,Tsujikawa:2015mga,Sola:2016zeg,Nesseris:2017vor,Kazantzidis:2018rnb} and Weak Lensing \cite{Hildebrandt:2016iqg,Kohlinger:2017sxk,Joudaki:2017zdt,Abbott:2017wau,Abbott:2018xao} cosmological data. Such measurements find a weaker growth rate of perturbations than anticipated in the context of the standard \lcdm model  \cite{Basilakos:2017rgc,Nesseris:2017vor,Kazantzidis:2018rnb,Macaulay:2013swa,Joudaki:2017zdt,Abbott:2017wau}. This weaker growth is expressed in the context of \lcdm parameters as a lower best fit value of the matter density parameter $\Omega_{0 \rm m} \approx  0.28 \pm 0.03$ \cite{Abbott:2017wau,Kazantzidis:2019dvk} than the one anticipated in the context of geometric probes including the CMB spectrum peak locations \cite{Ade:2015xua,Aghanim:2018eyx}  and the BAO data \cite{Aubourg:2014yra,Alam:2016hwk,Macaulay:2018fxi} in the context of flat \lcdm model ($\Omega_{0 \rm m}=0.315 \pm 0.007)$.

Other independent groups studying local expansion \cite{Rigault:2014kaa,Zhang:2017aqn,Dhawan:2017ywl,Fernandez-Arenas:2017isq,Freedman:2019jwv,Freedman:2020dne} find a lower value of $H_0$ compared to \cite{Riess:2019cxk} with larger errorbars reducing the effect of the tension to approximately $2 \sigma$. Moreover, independent measurements of the Hubble constant using $H(z)$ measurements \cite{Gott:2000mv,Chen:2011ab,Chen:2016uno,Yu:2017iju,Zhang:2018ida}, $\gamma-$rays \cite{Dominguez:2019jqc,Zeng:2019mae}, BAO measurements \cite{Wang:2017yfu}, as well as various combinations of data \cite{Lin:2017bhs,Abbott:2017smn,Haridasu:2018gqm,Park:2018tgj,Zhang:2018air,Ryan:2019uor,Cuceu:2019for} report a value for $H_0$ that is lower than the one provided by the local measurements and in consistency with the CMB measurement.

A wide range of models have been used to explain these tensions and properly extend \lcdm using specific new degrees of freedom (for a quantitative measure of tensions see Refs. \cite{Lin:2019zdn,Garcia-Quintero:2019cgt, Alexander:2019rsc}). For the Hubble tension these models include mechanisms that modify the scale of the sound horizon at last scattering using early dark energy \cite{Poulin:2018cxd,Karwal:2016vyq,Agrawal:2019lmo,Lin:2019qug, Braglia:2020bym, Smith:2019ihp} or other types of early species \cite{Bernal:2016gxb, Sakstein2019Nov, Ghosh:2019tab}, interacting dark energy with matter \cite{DiValentino:2019ffd,Yang:2018euj,DiValentino:2019jae,Yang:2018uae,Gomez-Valent:2020mqn, Vattis:2019efj}, screened fifth forces on the cosmic distance ladder \cite{Desmond:2019ygn,Desmond:2020wep}, modified gravity \cite{Kazantzidis:2020tko, Rossi:2019lgt, Ballardini:2016cvy, Braglia:2020iik, Lin:2018nxe}, local matter underdensities \cite{Lukovic:2019ryg} and new properties of late dark energy including new types of dark energy equation of state parameter \cite{Yang:2018qmz,Yang:2019jwn,Li:2019yem}. For the growth tension modified gravity \cite{DAmico:2016ntq,Nesseris:2017vor,Gonzalez-Espinoza:2018gyl,Kennedy:2018gtx,Linder:2018jil,Kazantzidis:2018rnb}, running vacuum models \cite{Sola:2017znb,Gomez-Valent:2018nib}, non-zero spatial curvature \cite{Ooba:2017ukj,Park:2017xbl} and modification of dark energy properties \cite{Pourtsidou:2016ico,Joudaki:2016kym,Melia:2016djn,Camera:2017tws,Gomez-Valent:2017idt,Ooba:2018dzf,Barros:2018efl,Lambiase:2018ows,Gomez-Valent:2018nib} have also been considered.

In both types of tension it has become clear that new properties of dark energy may constitute the required missing degree of freedom. In particular, it has been shown that a mildly phantom dark energy with equation of state parameter evolving slightly below $w=-1$ has the potential to resolve the Hubble tension by amplifying late time acceleration which leads to an increased best value of the Hubble parameter $H_0$ in the context of the CMB data, thus bringing it close to the value obtained by local distance ladder measurements \cite{Vagnozzi:2019ezj,DiValentino:2016hlg,Qing-Guo:2016ykt,DiValentino:2017zyq,Yang:2018qmz,Li:2019yem,Li:2019yem,Li:2020ybr}. Most previous analyses along the above lines utilize evolving equation of state parameters that in many cases have sophisticated functional forms. Even though such functional forms of $w(z)$, usually involve at most one new parameter, these approaches have two drawbacks: complexity of the $w(z)$ considered forms and worse fit than \lcdm to the Planck CMB TT power spectrum and other cosmological data ($\Delta \chi^2>0$). Thus, these models are usually not favoured \cite{Rezaei:2020mrj} compared to \lcdm in the context of information criteria that penalize models with additional parameters if they do not improve the quality of fit to data. It would therefore be desirable to construct models/parametrizations with no new parameters that can potentially resolve both the Hubble and growth tensions by modifying the dark energy properties. 

In particular, the following questions need to be addressed:
\begin{itemize}
\item
What are the properties of the new phantom degree of freedom required in order to increase the best fit value of $H_0$ in the context of CMB data to the level required for consistency with local measurements and resolution of the $H_0$ tension?
\item
What are the corresponding best fit values of cosmological parameters that emerge in the context this type of phantom dark energy and to what extend do they lead to improvement of the resolution of the growth tension?
\item
What is the quality of fit of these extended models to the CMB Planck and other cosmological data and how does it compare with the corresponding quality of fit of \lcdmnospace?
\end{itemize}
The goal of the present analysis is to address these questions using an approximate analytical method utilizing the degeneracies of cosmological parameters with respect to the form of the CMB power spectrum. In addition, we utilize more accurate numerical estimates of best fit cosmological parameters using Boltzmann and Markov Chain Monte Carlo (MCMC) codes. In the context of the analytical approximation we exploit the degeneracies of the CMB power spectrum among different cosmological parameter combinations and explore the consequences of variations of the dark energy equation of state parameter $w(z)$ on other cosmological parameters and in particular on the Hubble parameter $H_0$ and the matter density parameter $\Omega_{0 \rm m}$.

The structure of this paper is the following: In section \ref{secII} we review the well known degeneracies of the CMB TT power spectrum and identify the five cosmological parameter combinations that to a great extend uniquely determine the form of the spectrum. By demanding that these five combinations remain fixed to their \plcdm values, we identify the expected change of the best fit values of specific parameters including the Hubble parameter $H_0$ and the matter density parameter $\Omega_{0m}$ when the form of the dark energy equation of state parameter $w(z)$ changes. Thus we identify the forms of $w(z)$ leading to a value of $H_0$ consistent with local distance ladder measurements. In section \ref{secIII}, we fix $w(z)$ to the forms predicted analytically for the resolution of the $H_0$ tension and identify numerically the best fit cosmological parameters using Boltzmann and MCMC codes with Planck CMB data. We also compare the numerically obtained best fit cosmological parameter values with the corresponding values obtained in the context of the analytical approximation of section \ref{secII} for the same form of $w(z)$. Finally, in section \ref{secIV} we summarize and discuss possible extensions of this analysis.

\section{CMB spectrum degeneracies and the $H_0(w)$ dependence}\label{secII}
\begin{figure*}[ht!]
\centering
\includegraphics[width = 0.67 \textwidth]{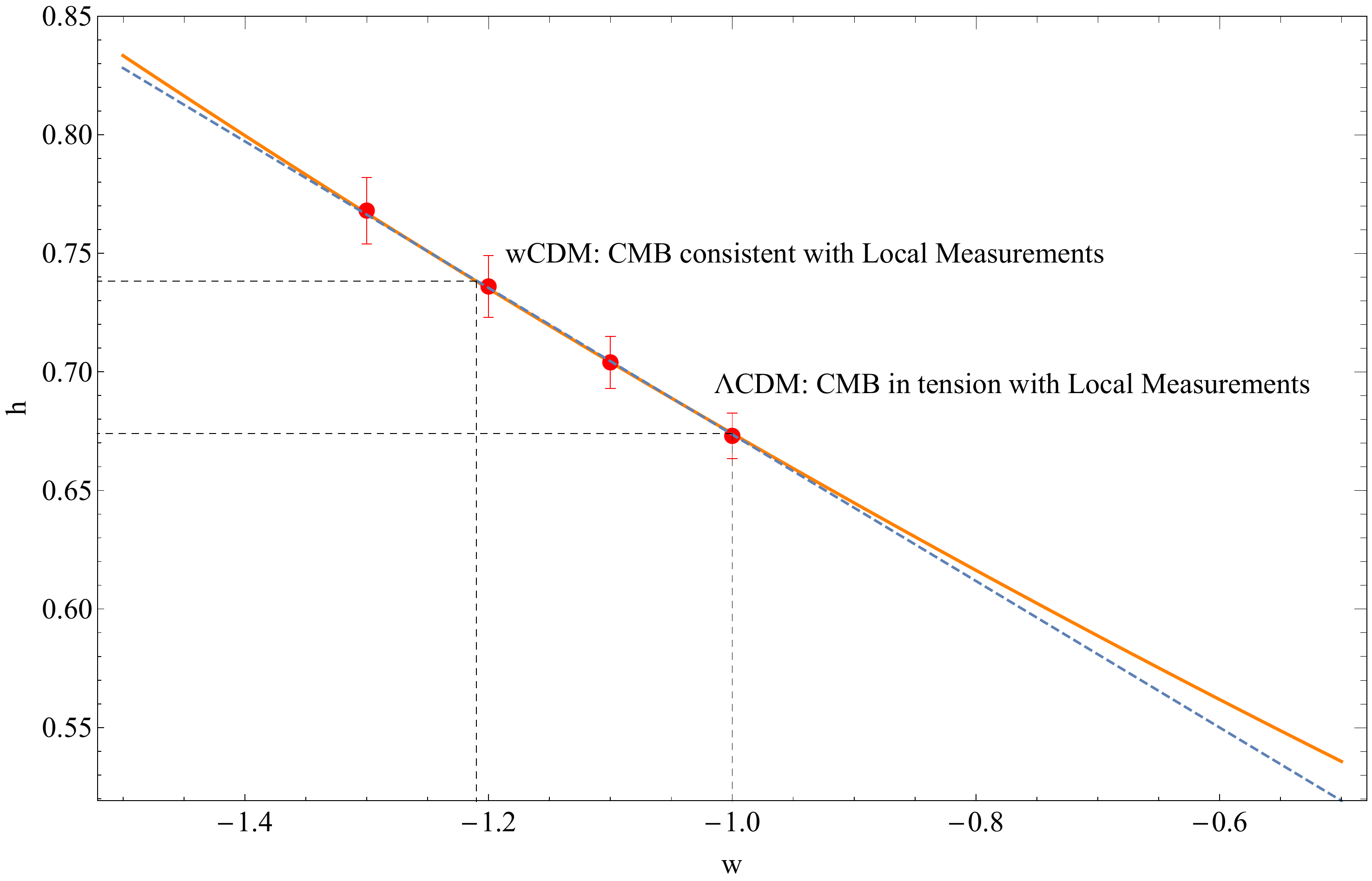}
\caption{The predicted value of $h$ as a function of the fixed $w$ for the one parameter dark energy ($w$CDM) model. The orange line corresponds to the theoretically predicted best fit values of $h$ for different values of $w$ in the case of the $w$CDM model, whereas the dashed blue line corresponds to the linear fitting that has been made. The red points display the actual best fit values, including the errorbars, of $h$ for specific values of $w$ obtained by fitting these models to the CMB TT anisotropy via the MGCosmoMC (see Table \ref{tab:homcosmo}).}
\label{fig:hwcdm}
\end{figure*} 

It is well known \cite{Efstathiou:1998xx,Elgaroy:2007bv} that the form of the CMB temperature power spectrum is almost uniquely determined, if the following parameter combinations are fixed
\begin{itemize}
\item
The matter density parameter combination $\omega_{\rm m}\equiv \Omega_{0 \rm m}h^2$ where $H_0=100 \; h\; km \; sec^{-1}\; Mpc^{-1}$.
\item
The baryon density parameter combination $\omega_{b}\equiv \Omega_{0 \rm b}h^2$  where $\Omega_{0 \rm b}$ is the present day baryon density parameter.
\item
The radiation density parameter combination $\omega_{\rm r}\equiv \Omega_{\rm 0r}h^2$  where $\Omega_{0 \rm r}$ is the present day radiation density parameter.
\item
The primordial fluctuation spectrum.
\item
The curvature parameter $\omega_{\rm k}=\Omega_{0 \rm k}h^2$.
\item
The flat universe co-moving angular diameter distance to the recombination surface
\be
d_A(\omega_m,\omega_r,\omega_b,h,w(z))=\int_0^{z_{r}}\frac{dz}{H(z)}
\label{dazr}
\ee
where $z_r\simeq 1100$ is the redshift of recombination provided to better accuracy as \cite{Hu:1995en}

\begin{eqnarray}
z_{\rm r} & = & 1048(1+0.00124\omega_{\rm b}^{-0.738})(1+g_1\omega_{\rm m}^{g_2})\label{zrfit}\\
g_1 & = & 0.0783\omega_{\rm b}^{-0.238}/(1+39.5\omega_{\rm b}^{0.763})\nonumber\\
g_2 & = & 0.560/(1+21.1\omega_{\rm b}^{1.81}).\nonumber
\end{eqnarray}
and $H(z)$ is the Hubble parameter at redshift $z$. The Hubble parameter takes the form
\begin{widetext}
\be
H(z,\omega_m,\omega_r,\omega_b,h,w(z))=H_0 \sqrt{\om0m (1+z)^3 + \Omega_{0 \rm r} (1+z)^4 + \Omega_{0 \rm de} e^{3\int_0^z dz'\; (1+w(z'))/(1+z')}}
\label{hz}
\ee
\end{widetext}
where $w(z)$ is the dark energy equation of state parameter at redshift $z$ and $\Omega_{0\rm de}=1-\Omega_{0 \rm m}-\Omega_{0 \rm r}$ is the present day value of the dark energy density parameter. The product $\sqrt{\omega_m} \cdot d_A$ is independent of $H_0$ and constitutes the well known shift parameter defined as  \cite{Efstathiou:1998xx,Wang:2006ts}
\be 
R= \sqrt{\omega_m} \int_0^{z_{r}}\frac{dz}{H(z)}
\ee
\end{itemize}

The observed values of the above parameter combinations as determined by the \plcdm CMB temperature power spectrum are the following \cite{Aghanim:2018eyx}
\bea
\bar \omega_m &=& 0.1430 \pm 0.0011
\label{planckomm}   \\
\bar\omega_b &=& 0.02237 \pm 0.00015 
\label{planckomb}  \\
\bar\omega_r&=& (4.64 \pm 0.3)\; 10^{-5}
\label{planckomr}  \\
\bar \omega_k &=& -0.0047 \pm 0.0029
\label{planckomk}   \\
\bar d_A &=& (100 \; km \; sec^{-1} \; Mpc^{-1})^{-1} (4.62\pm 0.08)
\label{planckda}
\eea

\noindent where for the radiation density we have assumed three relativistic neutrino species.

These parameter combinations also express the approximate degeneracy of the CMB with respect to various specific cosmological parameters. For example if the first four parameter combinations are fixed [eqs.  \eqref{planckomm} - \eqref{planckomk}], the fifth constraint [eq. \eqref{planckda}] provides the analytically predicted best fit value of the Hubble parameter $H_0$ (or $h$) given the dark energy equation of state parameter $w(w_0,w_1, ...,z)$  where $w_0,w_1, ...$ are the parameters entering the $w(z)$ parametrization\footnote{In the present analysis we assume a flat universe and fix $\bar \omega_k =0$.}. Thus, it is straightforward to use eqs. \eqref{dazr}, \eqref{hz}, \eqref{planckomm} and \eqref{planckda} to construct the function $h(w_0,w_1, ...)$ that gives semi-analytically the predicted best fit value of $h $ given a specific form of $w(z)$. This function is derived by solving the following equation with respect to $h$ 
\be
d_A({\bar \omega_m},{\bar \omega_r},{\bar \omega_b},h=0.674,w=-1)=d_A({\bar \omega_m},{\bar \omega_r},{\bar \omega_b},h,w(z))
\label{degeq}
\ee

\begin{figure*}[ht!]
\centering
\includegraphics[width = 0.8 \textwidth]{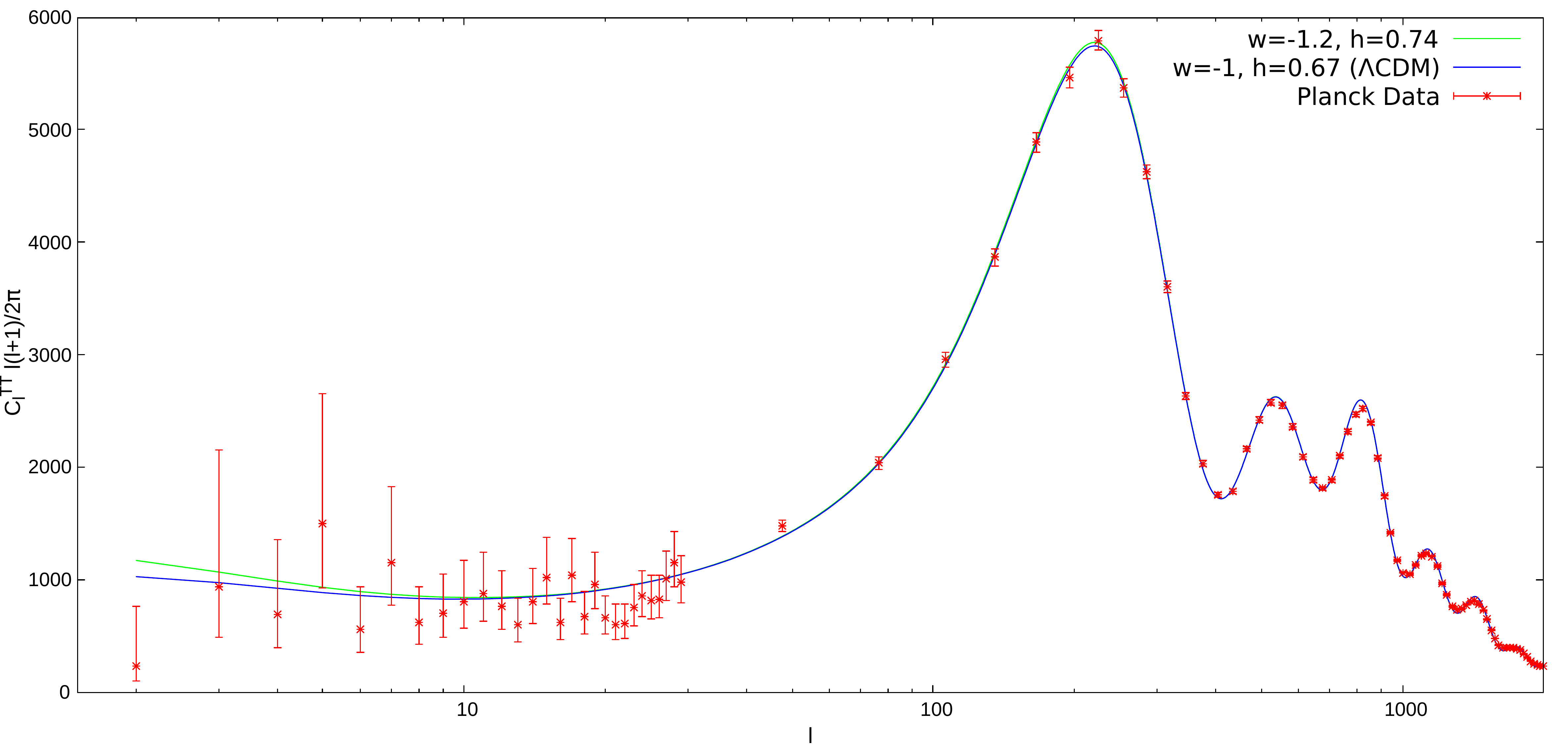}
\caption{The CMB power spectrum for \lcdm (blue line) and $w=-1.2$ (green line). We also show the binned high-$l$ and low-$l$ Planck data (red points).}
\label{fig:planck_spectrum}
\end{figure*} 

In the context of a simple one parameter parametrization where $w(z)$ remains constant in time and redshift, (\wcdm model), eq. \eqref{hz} takes the simple form 
\begin{widetext}
\be 
H(z,\omega_m,\omega_r,\omega_b,h,w(z))=H_0 \sqrt{\om0m (1+z)^3 +\Omega_{0 \rm r} (1+z)^4 + \left(1-\om0m -\Omega_{0 \rm r} \right) (1+z)^{3(1+w)}} \label{hzwcdm}
\ee
\end{widetext}
and using the above described approach, solving eq. (\ref{degeq}) it is straightforward to derive the degeneracy function $h(w)$ shown in Fig. \ref{fig:hwcdm} (continuous orange line). In the range $w\in [-1.5,-1]$, $h(w)$ is approximated as a straight line (dashed blue line in Fig. \ref{fig:hwcdm})
\be
h(w)  \approx -0.3093w + 0.3647
\label{hweq}
\ee 

The points with the errorbars were obtained by fitting to the Planck/CMB power spectrum using the corresponding $w$CDM models with  fixed $w$. This analysis is discussed in more detail in the next section. In Fig. \ref{fig:planck_spectrum} we show the predicted form of the CMB TT anisotropy spectrum for $w=-1$ ($h=0.67$, $\Omega_{0m}=0.314$) and $w=-1.2$ ($h=0.74$, $\Omega_{0m}=0.263$)  demonstrating the invariance of the CMB power spectrum when the cosmological parameters are varied along the above described degeneracy directions.

The dark energy equation of state parameter value leading to $h(w)=0.74$ is $w \approx -1.217$ which is  the predicted value of $w$ required to alleviate the $H_0$ tension, a result consistent with previous studies \cite{Vagnozzi:2019ezj, DiValentino:2016hlg}. In particular, a related analysis has been performed in \cite{Vagnozzi:2019ezj}, where the author points out that fixing the dark energy equation of state $w \approx -1.3$ or the effective number of relativistic species $N_{eff} \approx 3.95$ may lead to the relaxation of the $H_0$ tension. The novel feature of our work is the use of analytical methods to identify the qualitative features required for any form of $w(z)$ to relax the $H_0$ tension. 

This method for deriving the predicted dark energy properties required to resolve the $H_0$ tension may be extended to more parametrizations of $w(z)$. For example in the case of the two parameter CPL parametrization \cite{Chevallier:2000qy,Linder:2002et} expansion of $w(z)$
\be
w=w_0 + w_1 (1-a)= w_0 + w_1 z/(1+z)
\label{cpl}
\ee
eq. \eqref{hz} is written as 
\begin{widetext}
\be
H(z)=H_0 \sqrt{\om0m (1+z)^3 +\Omega_{0 \rm r} (1+z)^4 +\left(1-\om0m - \Omega_{0 \rm r}\right) 
(1+z)^{3(1+w_0+w_1)}  e^{-3 \frac{w_1 z}{1+z}}} \label{hzcpl}
 \ee
\end{widetext}

\begin{figure}[h!]
\centering
\includegraphics[width = 0.45\textwidth]{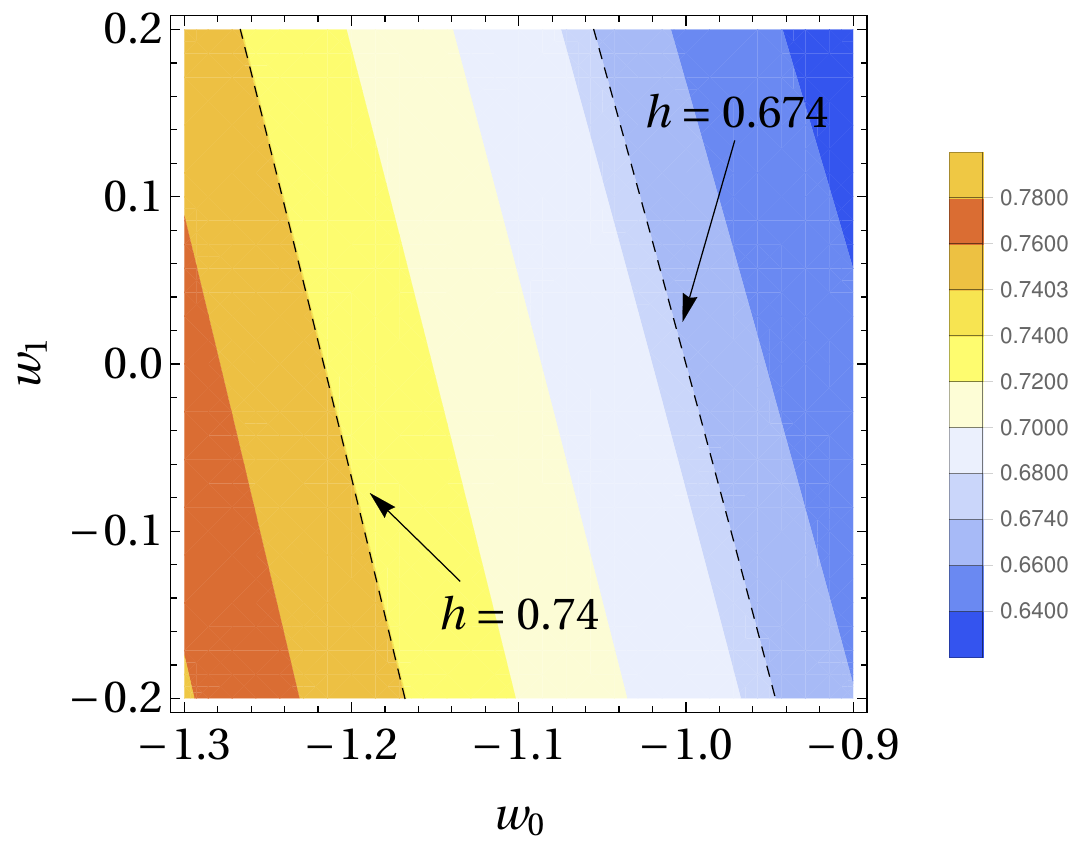}
\caption{The degeneracy with respect to the CMB spectrum in the parameter space $(w_0-w_1)$. The dashed lines correspond to $h=0.674$ (\lcdm value) and to $h=0.74$ (the value of Ref. \cite{Riess:2019cxk}).}
\label{fig:cpl_contour}
\end{figure} 

\begin{figure}[h!]
\centering
\includegraphics[width = 0.45\textwidth]{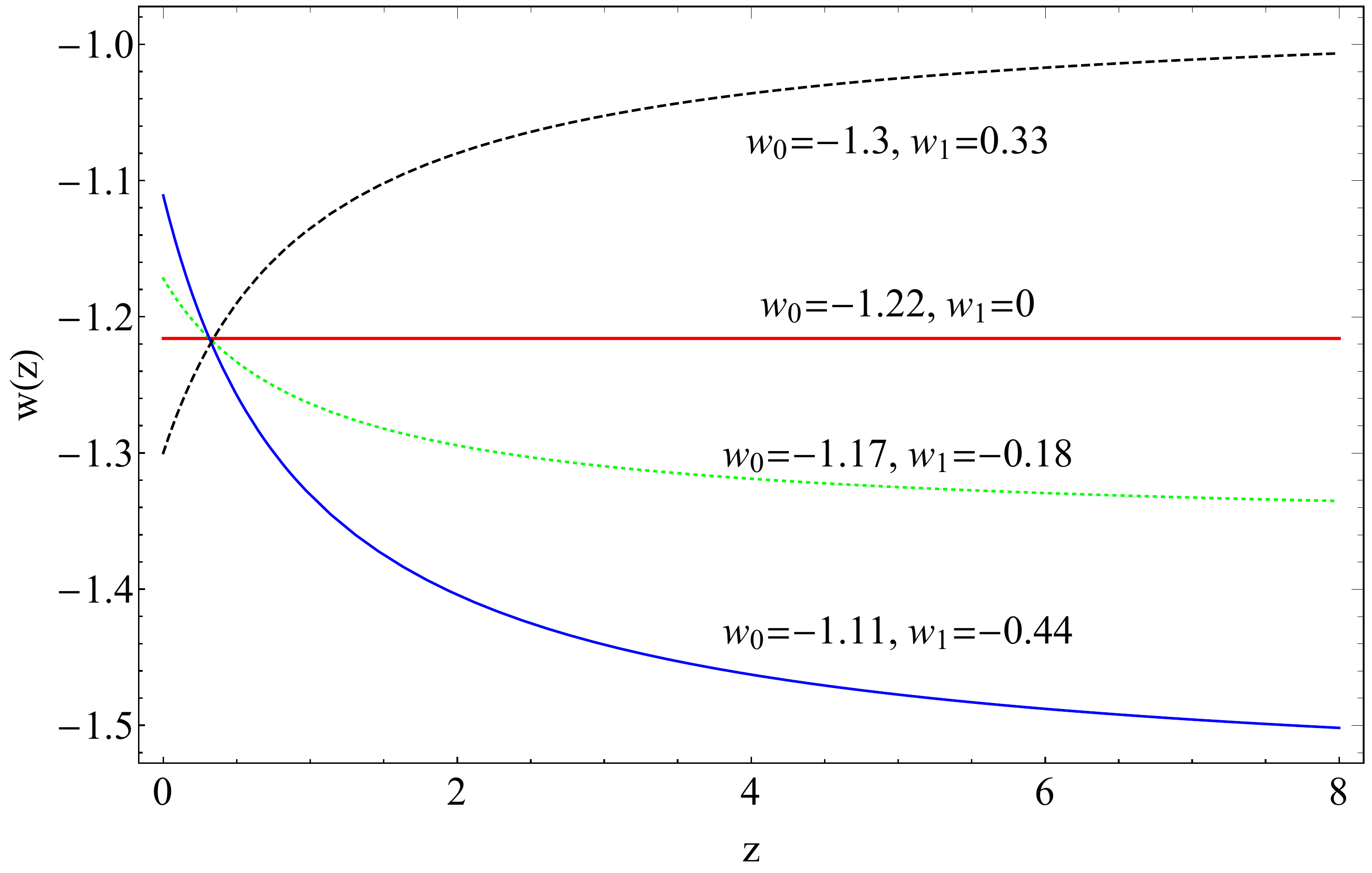}
\caption{The evolution of $w(z)$ for various values of $(w_0,w_1)$ along the degeneracy $h=0.74$ line of Fig.  \ref{fig:cpl_contour}. All these parameter values lead to a best fit value $h=0.74$ in the context of the CMB power spectrum. However, they do not have the same quality of fit to other cosmological data which can be used to break this model degeneracy. The common $(z, w)$ point of intersection of all the $w(z)$ plots is $(0.31, -1.22)$.}
\label{fig:wcpl_line}
\end{figure} 

Using now eqs. \eqref{dazr}, \eqref{planckomm}, \eqref{planckda} and \eqref{hweq} in the context of the above described method, it is straightforward to derive the degeneracy function $h(w_0,w_1)$, by solving eq. (\ref{degeq}). This is shown in Fig. \ref{fig:cpl_contour}. The dashed lines correspond to the parameter values that satisfy $h(w_0,w_1)=0.674$ [the \lcdm value which as expected goes through the point $(w_0,w_1)=(-1,0)$] and   $h(w_0,w_1)=0.74$ (the local distance ladder measurements  value). The constant $h$ lines shown in Fig. \ref{fig:cpl_contour} are approximately straight lines in the range of the  $w_0-w_1$ parameter space shown in Fig. \ref{fig:cpl_contour}. In particular, for the case of the value $h = 0.74$, which alleviates the $H_{0}$ tension, this line is approximated by the equation,
\be
w_1 \approx  -4.17 w_0 - 5.08
\label{w0w1deg}
\ee
Clearly, the preference for a phantom like behaviour $w(z)<-1$ in the context of the local measurement value of $h$,  at least for some redshift range is apparent in Fig. \ref{fig:cpl_contour}. This is also demonstrated in Fig. \ref{fig:wcpl_line} where we show four forms of $w(z)$ based on the CPL parametrization that can resolve the $H_0$ tension by providing a best fit value of $h=0.74$ from the CMB data.  The corresponding $w$CDM value of $w=-1.22$ is also shown. Clearly all degenerate forms of CPL $w(z)$ that relax the $H_0$ tension go through the same point at $z=0.31$ crossing the $w=-1.22$ line. This type of degeneracy in particular redshifts for cosmological parameters has been discussed in Ref. \cite{Kazantzidis:2018jtb}. Also degenerate $w(z)$ curves with $w_0<1.22$ are increasing functions of $z$, while those with $w_0>1.22$ are decreasing functions of $z$. This appears to be a general feature of all $w(z)$ parametrizations that can relax the $H_0$ tension. For example the PEDE parametrization \cite{Li:2019yem} and the late dark energy transition hypothesis \cite{Benevento2020Feb} with $w(z\simeq 0)>-1.22$ are decreasing functions of the redshift $z$ as predicted by the above degeneracy analysis. The identification of these properties opens up the possibility of a very late type phase transition at $z\simeq 0.01$ from a phantom phase to a \lcdm phase with a sharply increasing rather than decreasing function of $w(z)$.

Even though the approximate parameter degeneracy exploited in this section is useful for the derivation of the forms of $w(z)$ that can alleviate the $H_0$ tension, an important fact that needs to be considered is the quality of fit of the preferred degenerate forms of  $w(z)$ to other cosmological data like  SnIa, BAO and growth of perturbations data (Redshift Space Distortion $f\sigma_8(z)$ and weak lensing data) as well as to actual CMB power spectrum data which may not fully respect the above exploited approximate degeneracy (especially at low $l$). Such a fit to cosmological data beyond the CMB is expected to break the above degeneracy obtained from the CMB spectrum. Even if particular forms of $w(z)$ can lead to apparent alleviation of the $H_0$ tension such a solution would not be preferable if the quality of fit to the actual CMB spectrum and to other cosmological data is significantly degraded compared to \lcdm ($w=-1$). Thus, in the next section we address the following questions:
\begin{itemize}
\item
What is the quality of fit of the forms of $w(z)$ that are predicted to resolve the $H_0$ tension, on cosmological data involving SnIa, BAO, growth Redshift Space Distortion data and the actual Planck CMB TT power spectrum data?  Is this quality of fit $(\chi^2)$ similar to the corresponding quality for \lcdmnospace?
\item
Is the $H_0$ tension actually alleviated when the full CMB spectrum data are used in the context of a model with fixed $w(z)$ to its predicted form (\eg $w=-1.22$ in the context of a constant $w$)? 
\item
Is the growth tension partially relaxed in the context of the above preferred $w(z)$ found?
\end{itemize}
These questions will be addressed mainly in the context of a redshift independent $w$ but it is straightforward to generalize the analysis for more general forms of $w(z)$.

\section{Numerical analysis of dark energy models}\label{secIII}
\begin{figure*}[ht!]
\centering
\includegraphics[width = 0.8 \textwidth]{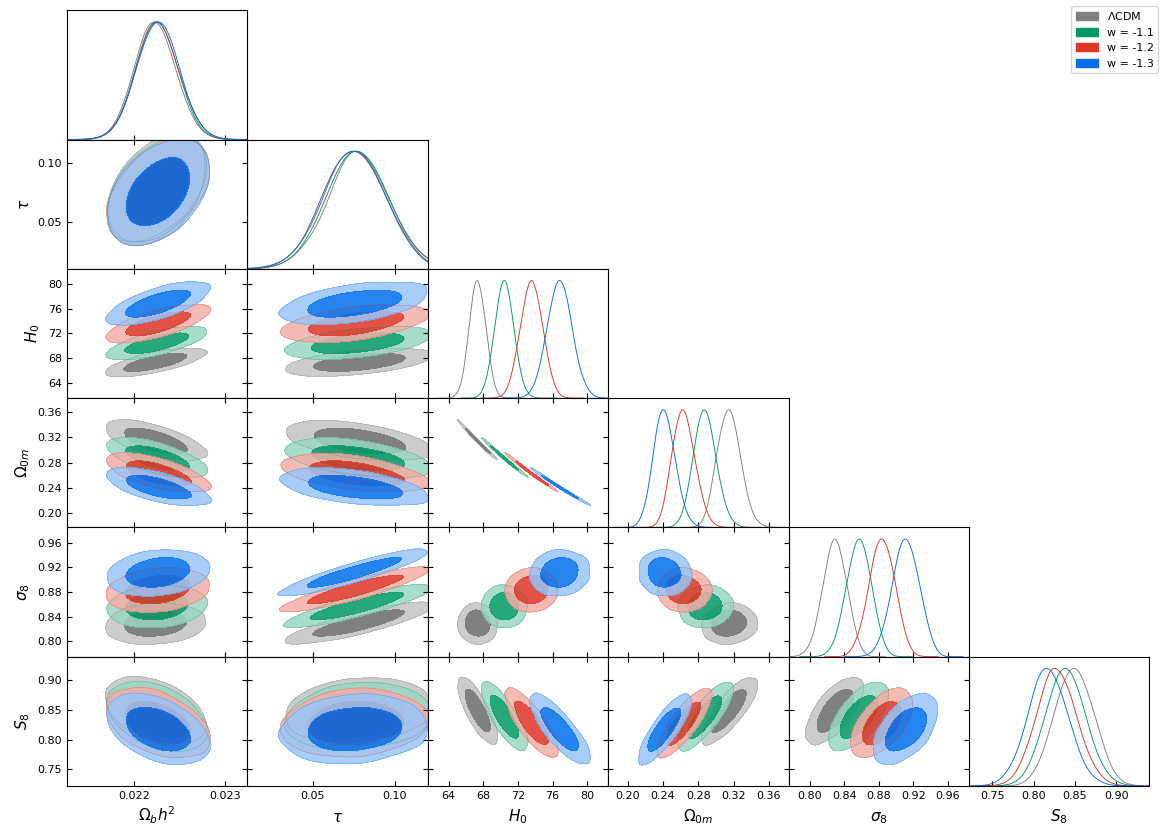}
\caption{The contour plots constructed with MGCosmoMC using the \textit{Planck}TT and lowP likelihoods for \lcdm and $w$CDM models. The gray contours correspond to the \lcdm model. The green contours correspond to $w = -1.1$, the red ones to $w = -1.2$, while the blue to $w=-1.3$.  For $w=-1.1$, the best fit value of $H_0$ is close to that of the \plcdm measurement \cite{Ade:2015xua}, while the $w=-1.2$  and $w=-1.3$ values shift $h$ closer to the local distance ladder measurements \cite{Riess:2019cxk}.}
\label{fig:contours}
\end{figure*} 

In order to test the resolution of the $H_0$ tension and test the quality of fit to the CMB and other cosmological data of the models discussed in the previous section, we use the MGCosmoMC numerical package \cite{Zhao:2008bn,Hojjati:2011ix,Zucca:2019xhg} with the \textit{Planck} dataset. In particular, we use the \textit{Planck} TT and lowP dataset, \ie the TT likelihood for high-l multipoles $(l>30)$ as well as the Planck temperature and polarization data for low multipoles $(l<30)$. The priors that have been used as input can be seen in Table \ref{tab:priors}.

\begin{table}[h!]
\caption{The MGCosmoMC priors that have been used in Figs. \ref{fig:contours} and Fig. \ref{fig:contourstot}. We also set $A_{lens} = 1$ and $\Omega_{k} = 0$.}
\label{tab:priors}
\begin{centering}
\begin{tabular}{|c|c|}
 \hline 
 \rule{0pt}{3ex}  
Parameters  & Priors\\
\hline
\rule{0pt}{3ex}  
$\Omega_{b}h^{2}$  & $[0.005, 0.1]$\\
\hline
\rule{0pt}{3ex}  
$\Omega_{c}h^{2}$  & $[0.001, 0.99]$\\
\hline
\rule{0pt}{3ex}    
$100 \theta_{MC}$  & $[0.5, 10]$\\
\hline  
\rule{0pt}{3ex}  
$\tau$  & $[0.06, 0.8]$\\
\hline  
\rule{0pt}{3ex}  
$ln \left(10^{10} A_s \right)$  & $[1.61, 3.91]$\\
\hline
\rule{0pt}{3ex}  
$n_{s}$  & $[0.8, 1.2]$\\
\hline
\end{tabular}
\end{centering}
\end{table}

We  fix $w$ to the values of the points shown in Fig. \ref{fig:hwcdm} ($w=-1.0, -1.1, -1.2, -1.3$)  and construct  the likelihood contours for the cosmological parameters of these four models. The resulting best fit values of $h$ are shown in Table \ref{tab:homcosmo} (see also Fig. \ref{fig:hwcdm}) and are in excellent agreement with the expectations based on the parameter degeneracy analysis of the previous section (orange continuous line in Fig. \ref{fig:hwcdm}). The corresponding likelihood contours are shown in Fig. \ref{fig:contours}.

\begin{table*}[ht!]
\caption{The analytically predicted CMB best fit values of $h$ and $\Omega_{0m}$ for fixed $w$, obtained by using the CMB parameter degeneracy arguments, as well as the ones obtained by the actual fit of the corresponding $w$ model to the Planck TT CMB anisotropy power spectrum. The quality of fit for each model compared to \lcdm is also indicated by the value of $\Delta \chi^2$.}
\label{tab:homcosmo}
\begin{centering}
\begin{tabular}{|c|c|c|c|c|c|c|}
 \hline 
 \rule{0pt}{3ex}  
$w$  & $\Omega_{0m}^{th}$ & $h_{th}$ & $\Omega_{0m}^{obs}$ & $h_{obs}$ & $\chi^2_{CMB}$ & $\Delta \chi^2_{CMB}$\\
    \hline
    \rule{0pt}{3ex}  
$-1.0$ & $0.316$ & $0.674$ & $0.315\pm 0.013$ & $0.673\pm 0.010$ & $11266.516$ &$-$\\
$-1.1$ & $0.289$ & $0.704$ & $0.288\pm 0.013$ & $0.704\pm 0.011$ & $11266.530$ &$0.014$\\
$-1.2$ & $0.265$ & $0.735$ & $0.263^{+0.012}_{-0.014}$ & $0.736\pm 0.013$ & $11267.132$ &$0.616$\\
$-1.3$ & $0.244$ & $0.766$ & $0.242^{+0.012}_{-0.013}$ & $0.768\pm 0.014$ & $11266.520$ &$0.004$\\
\hline
\end{tabular}
\end{centering}
\end{table*}

\begin{figure*}[ht!]
\centering
\includegraphics[width = 0.8 \textwidth]{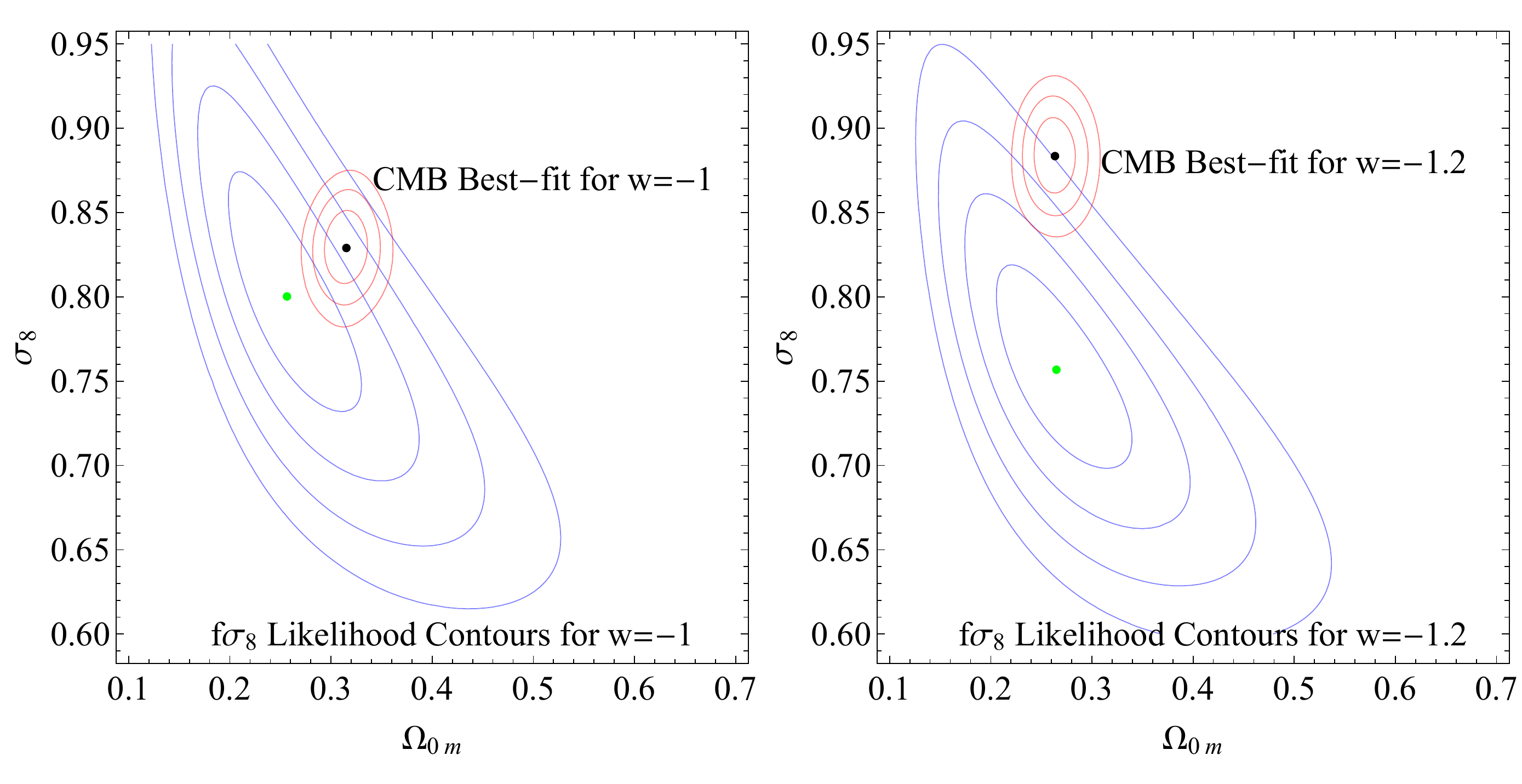}
\caption{The $1\sigma-4\sigma$ contours in the parametric space $\Omega_{0 \rm m}-\sigma_8$. The blue contours correspond to the
best fit growth compilation of Ref. \cite{Sagredo:2018ahx}, while the red to the $1\sigma-4\sigma$ confidence contours for $w=-1$ (left panel) and $w=-1.2$ (right panel) obtained from the Planck data.}
\label{fig:Growth_CMB}
\end{figure*} 

Clearly, the likelihood contours for the Hubble parameter shift to higher best fit values as $w$ decreases in the phantom regime ($w<-1$). At the same time the best fit values of the matter density parameter $\Omega_{0 \rm m}$ decrease in accordance with the degenerate parameter combination $\Omega_{0 \rm m} h^2$. 

This reduced value of the best fit  $\Omega_{0 \rm m}$ would naively imply reduced growth of cosmological perturbations and thus resolution of the growth tension.  However, the reduced best fit value of the matter density parameter $\Omega_{0 \rm m}$ matter density is not enough to soften the growth tension, since the best fit value of the  parameter $\sigma_8$ (the present day rms matter fluctuations variance on scales of $8 h^{-1} Mpc$) appears to increase more rapidly, as $w$ decreases in the phantom regime. Since this parameter is proportional to the initial amplitude of the matter perturbations power spectrum, its increase amplifies the growth of perturbations and tends to cancel the effect of the decrease of the best fit $\Omega_{0 \rm m}$ in the context of perturbations growth. This is demonstrated in Fig. \ref{fig:Growth_CMB} where we show the $\sigma_8$ likelihood contours obtained by fitting the models $w=-1$ (\lcdmnospace) and $w=-1.2$ to the growth $f\sigma_8$ data (we have used the conservative robust dataset of Table 2 of Ref. \cite{Sagredo:2018ahx}, a subset of an up to date compilation presented in \cite{Skara:2019usd}). Superimposed we also show the corresponding likelihood contours obtained from the Planck CMB TT power spectrum obtained for each value of fixed $w$. Clearly, the tension between the RSD $f\sigma_8$ data and the Planck data increases in the context of the phantom model $w=-1.2$ compared to \lcdm ($w=-1$).

\begin{figure*}[ht!]
\centering
\includegraphics[width = 0.8 \textwidth]{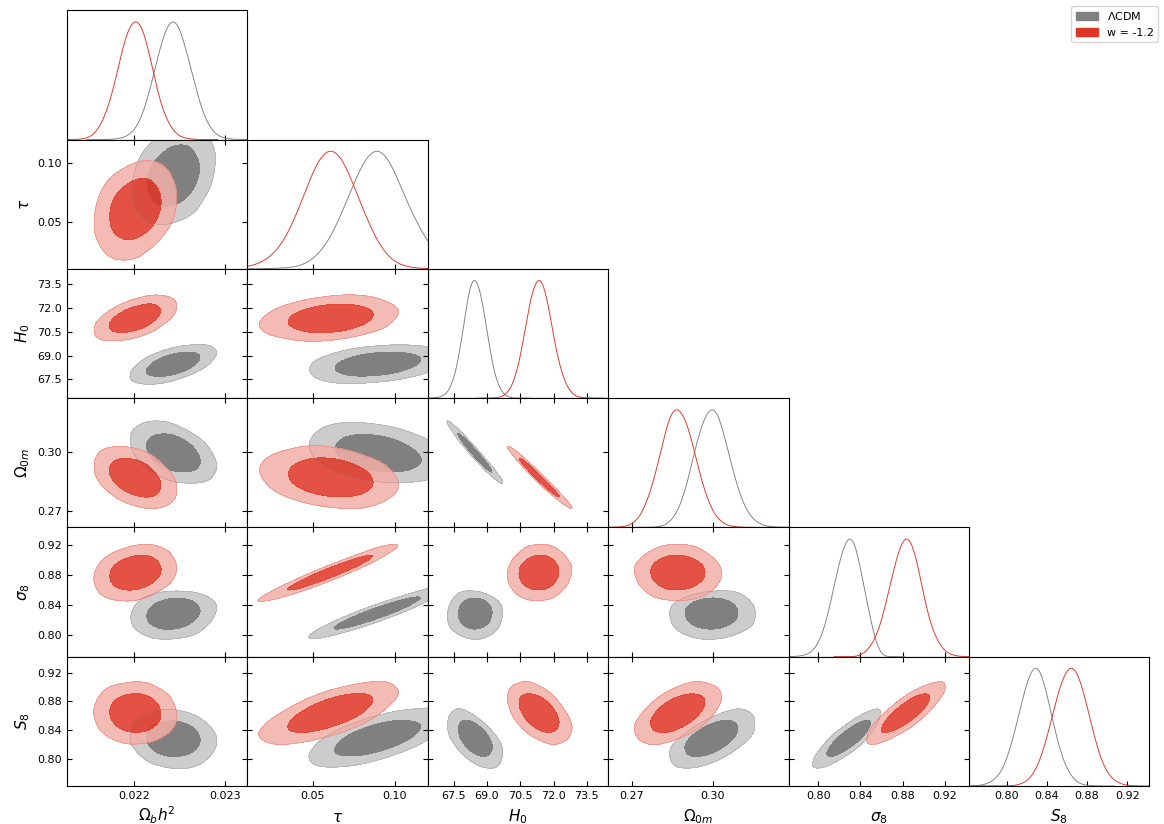}
\caption{The likelihood contours constructed with MGCosmoMC using the cosmological data combination of Pantheon SnIa \cite{Scolnic:2017caz}, BAO data \cite{Beutler:2011hx,Ross:2014qpa,Alam:2016hwk}, CMB data \cite{Ade:2015xua}, as well as the prior of the Hubble constant \cite{Riess:2019cxk} for \lcdm (gray contours) and $w$CDM with $w=-1.2$ (red contours).}
\label{fig:contourstot}
\end{figure*}

In addition to the growth data we also fit the models $w=-1$ and $w=-1.2$ to a cosmological data combination including the Pantheon SnIa \cite{Scolnic:2017caz}, BAO data \cite{Beutler:2011hx,Ross:2014qpa,Alam:2016hwk}, CMB data \cite{Ade:2015xua}, as well as the prior of the Hubble constant published by Riess et al. \cite{Riess:2019cxk} and obtain for \lcdm $\chi^2=12319.2$, while for $w=-1.2$ we obtain $\chi^2=12332.7$. We thus find $\Delta \chi^2=13.5$. This difference of $\Delta \chi^2 =13.5$  for the phantom model, indicates a significantly reduced quality of fit compared to \lcdm in agreement with previous studies \cite{Arendse:2019hev}. The corresponding likelihood contours are shown in Fig. \ref{fig:contourstot}. It is therefore clear that the particular fixed $w$ models considered here lead to an apparent resolution of the Hubble tension since they increase the best fit value of $H_0$ in the context of the CMB data but the resolution is not viable since the growth tension gets worse while the quality of fit of these models  to the SnIa and BAO data is not as good as for \lcdmnospace. This result is consistent with previous studies \cite{DiValentino:2016hlg}, where it has been demonstrated that non-CMB data, such as BAO and SNIa favour lower values of $H_0$ which are more consistent with the CMB value, while also disfavouring $w<-1$ in the context of flat and non-flat untilted inflation models \cite{Park:2018tgj}. It is, however, worth mentioning that for the combination of the CMB Planck data and the Riess Hubble constant prior the quality of the fit improves drastically for $w = -1.2$, with $\Delta\chi^2 = -10.7$ in respect to $w = -1$. The exploitation of the CMB spectrum degeneracy of more complicated forms of $w(z)$ however may lead to better fits to growth, SnIa and BAO cosmological data. 

\section{Conclusion - Discussion - Outlook}\label{secIV}

We have used analytical degeneracy relations among cosmological parameters and numerical fits to cosmological data to identify the qualitative and quantitative features of dark energy models that have the potential to relax the $H_0$ tension of the \lcdm model. We have found that mildly phantom models with mean equation of state parameter $w\simeq -1.2$ have the potential to alleviate this tension. The models may be constructed in such a way that there are no extra parameters compared to \lcdm by using fixed parametrizations of $w(z)$. In practice however these  models involve more fine tuning compared to \lcdm and are clearly less natural than the standard model. In addition the quality of fit of the simplest of such models to cosmological data beyond the CMB is not as good as the corresponding quality of fit of \lcdmnospace. However, it is straightforward to construct physical models involving either phantom scalar field with non-canonical kinetic terms or modified gravity models that naturally produce the required phantom behavior of dark energy. Despite the usual stability issues of such models it is possible to construct ghost free versions \cite{Perivolaropoulos:2005yv}. For example, physical models described by scalar field Lagrangians can reproduce an effective dark energy with a constant equation of state parameter $w$ in the context of both quintessence ($w>-1$) \cite{Peebles:1987ek,Ratra:1987rm,Zlatev:1998tr} and phantom dark energy $w<-1$ \cite{Chiba:1999ka}.

In particular, a dynamical dark energy scalar field with an inverse power law potential of the form $V(\phi)=M^{(4+\alpha)} \, \phi^{-\alpha}$  (where $M$ and $\alpha>0$ are free parameters), corresponds to a physically interesting model where the dark energy equation of state parameter $w$ is constant and takes the form \cite{Zlatev:1998tr},
\be
w = \frac{\frac{\alpha}{2}w_B - 1}{1 + \frac{\alpha}{2}}\label{wStein}
\ee  
where $w_B$ is the equation of state parameter of the dominant background. Clearly, for a matter dominated epoch $(w_B=0)$, and $\alpha>0$, we can obtain a constant $w$ and a quintessence like behaviour $(w>-1)$.

Similarly, a phantom like behaviour $(w<-1)$ with constant $w$, may be obtained \cite{Chiba:1999ka} in the context of a scalar field with non-canonical kinetic terms with an action of the form
\be 
S = \int d^4x\sqrt{-g}\left({\frac{1}{2\kappa ^2}}R+p(\phi,\nabla
  \phi)\right)+S_B
\ee
where $\kappa^2= 8 \pi G$ and $S_B$ is the action of the background. The  Lagrangian may be assumed to depend only on the scalar
field $\phi$ and its derivative squared $X=-\frac{1}{2} \nabla^\mu \phi \nabla_\mu \phi$. In the case of a slowly varying field $X$ the pressure $p$ and energy density $\rho$ of the field take the form \cite{Chiba:1999ka}
\begin{align}
&p= f(\phi)(-X+X^2),
\label{eq:p_q} \\
&\rho = 2X{\partial p\over \partial X}-p= f(\phi)(-X+3X^2)
\label{eq:rho_q}
\end{align}
For $f(\phi) \propto \phi^{- \alpha}$, eqs. \eqref{eq:p_q} and \eqref{eq:rho_q} lead to  an equation of state parameter of the form 
\be 
w = \frac{(1+w_B)\alpha}{2}-1.\label{wchiba}
\ee 
For a matter dominated epoch ($w_B=0$) an appropriate value of $\alpha$  can lead to either a quintessence or a phantom behavior. In particular for $\alpha<2$ we obtain $w>-1$ (quintessence behavior), while for $a<0$ we obtain a physical model with $w<-1$  (phantom equation of state).

However, the constant $w$ behavior of both of the above physical models described by eqs. \eqref{wStein} and \eqref{wchiba} is a good approximation only in the context of a dominant background fluid with constant equation of state $w_B$. In our universe this would occur for example only well in the matter dominated epoch.  These equation of state parameters would cease to have a constant form near the end of the matter era and in the present transition cosmological era. Thus, the constancy of $w$ in the context of these physical models is a good approximation only on high redshifts ($z>2$). 

Interesting extensions of the present analysis include the following:
\begin{itemize}
\item
A comparative analysis of phantom models identified using the degeneracy analytical method proposed here involving also redshift dependence of $w(z)$. Such an analysis would rank these models according to their quality of fit on cosmological data. 
\item
The construction of physical models that can reproduce the forms of $w(z)$ required to relax the $H_0$ and possibly the growth tension as well, while providing a better fit to the cosmological data than the fit of \lcdmnospace. The construction of stable theories with phantom behavior is possible in the context of modified gravity theories. In many such theories however  including $f(R)$ and  scalar-tensor theories, it is not possible to combine stability, with the weaker gravity and phantom behaviour  \cite{Polarski:2016ieb,Gannouji:2018ncm} required for the resolution of the $H_0$ and growth tensions.
\end{itemize}

The analytical approach for the $H_0$-$w(z)$ degeneracy pointed out in the present analysis offers a new method to systematically search and design $w(z)$ forms that can combine the proper features required to consistently relax the tension while keeping a good fit to other cosmological data. Our goal  here was only to introduce the method and apply it to the simplest cases while also pointing out the difficulties in resolving the tension. In a subsequent full application and extension of the method we plan to exploit its full potential in identifying possible forms of $w(z)$ that can actually resolve the tension while keeping good fit to other cosmological data.

\textbf{Numerical Analysis Files}: The numerical files for the reproduction of the figures can be found in \cite{numcodes}.

\section*{Acknowledgements}
We thank Savvas Nesseris for his help with the MGCosmoMC chains. All the runs were performed in the cluster of the Institute of Theoretical Physics (IFT) in Madrid. This research is co-financed by Greece and the European Union (European Social Fund- ESF) through the Operational Programme ``Human Resources Development, Education and Lifelong Learning" in the context of the project ``Strengthening Human Resources Research Potential via Doctorate Research – 2nd Cycle" (MIS-5000432), implemented by the State Scholarships Foundation (IKY).

\raggedleft
\bibliography{Bibliography}

\end{document}